\documentstyle[12pt,epsfig]{article}
\begin{document}
\setlength{\baselineskip}{0.65cm}
\setlength{\parskip}{0.5cm}
\begin{titlepage}
\begin{flushright}
DESY 95--068 \\ April 1995
\end{flushright}
\vspace{1.5cm}
\begin{center}
\Large
{\bf Dynamical Parton Distributions}\\
\vspace{0.1cm}
{\bf for $ \alpha_{S} $ Determinations} \\
\large
\vspace{2.5cm}
A. Vogt$\, ^{\ast}$ \\
\vspace{1.5cm}
\normalsize
{ Deutsches Elektronen-Synchrotron DESY \\
 Notkestra{\ss}e 85, D-22603 Hamburg, Germany} \\
\vspace{4.0cm}
\large
{\bf Abstract} \\
\end{center}
\vspace{-0.3cm}
\normalsize
Precise determinations of the strong coupling constant in hadronic
collisions demand sets of parton densities spanning a sufficiently
wide range of values for the QCD scale parameter $\Lambda $. For such
applications, we supplement the recent GRV parametrization by five
sets of dynamically generated parton distributions for $\Lambda
$-values corresponding to $ \alpha _{S} (M_{Z}^{2}) $ between 0.104
and 0.122. Present DIS data, including recent small-$ x $ HERA results,
do not appreciably discriminate between these sets.
\vfill
\noindent
\small
$^{\ast} $ On leave of absence from Sektion Physik, Universit\"at
M\"unchen, D-80333 Munich, Germany
\normalsize
\end{titlepage}

Parton distributions are essential for connecting observables measured
in hard hadronic collisions to the underlying partonic interactions.
Parametrizations of these quark and gluon distributions always include
a specific value of the QCD scale parameter $ \Lambda $ via the
appearance of this quantity in the renormalization group evolution
equations. This simple fact leads to a question of consistency \cite
{Web} when such parametrizations are employed in extractions of the
strong coupling constant $ \alpha _{S}(Q^{2}) $ from data on, e.g.,
jet production rates at the DESY $ ep $ collider HERA \cite{HEa} or on
$ W + \,$jet production at the Fermilab TEVATRON \cite{FLa}. Hence
precise analyses of this type demand sets of parton densities spanning
a sufficiently wide range of values for $\Lambda $.
Recent parametrizations of CTEQ \cite{CTEQ}, MRS \cite{MRS,MRSA} and GRV
\cite{GRV92,GRV94}, however, have been provided only for some `central'
values, $ \Lambda_{\overline{MS}}^{(4)} = 200 \ldots 250 \mbox{ MeV} $,
of the NLO ($\overline{\mbox{MS}} $) scale with four active quarks
flavours. The only exception is CTEQ2 \cite {CTEQ}, where one additional
large-$ \Lambda $ set has been included.

In this paper, we present five parton density sets which cover,
together with the recent GRV parametrization \cite{GRV94}, the range of
$ \Lambda_{\overline{MS}}^{(4)} $ from 150 to 400 MeV in steps of 50
MeV, corresponding to $ \alpha _{S}(M_{Z}^{2}) $ between 0.104 and
0.122. For the derivation of these sets we follow the approach
developed in \cite{GRV92}, in which the quark and gluon densities are
generated from valence-like inputs at a very low resolution scale
$ Q^{2} = \mu ^{2} $. Within their uncertainties due to the precise
choice of $ \mu ^{2} $ \cite{GRV93}, those `dynamical' distributions
quantitatively anticipated the strong small-$ x $ rise of the proton
structure function $ F_{2}^{p} $ recently observed at HERA \cite
{ZEUSf2,H1f2}. Concerning the  present task, it should be noted that
this framework provides a close connection between the experimentally
tightly constrained quark densities and the much less well-known gluon
distribution. This connection leads to parametrizations which are as
similar as possible for different values of $ \Lambda $ and thus are
very well suited for studying the quantitative importance of the
consistency question raised above.

Technically, the present paper closely follows the recent GRV update
\cite{GRV94} wherever possible. Our analysis is carried out in the next
to leading order (NLO) perturbative QCD framework \cite{FP}. We employ
the $\overline {\mbox{MS}} $ renormalization and mass factorization
schemes with a `running' number $ f $ of active flavours in the QCD
beta function and a fixed number of three light partonic quark species
present in the $Q^{2}$-evolution \cite{GRS}. The NLO relation connecting
$ \alpha_{S}(Q^{2}) $ and $ \Lambda_{\overline{MS}} $ is chosen as in
eq.\ (10) of \cite{GRV94}, and the heavy quark thresholds in the beta
function are taken to be $ Q_{h} = m_{h} = 1.5\,(4.5) \mbox{ GeV} $ for
the charm (bottom) case, respectively. The values of $ \Lambda^{(f)}_
{\overline{MS}} $, connected by requiring the continuity of $ \alpha_{S}
$ at these thresholds, are listed in Table 1 below.
The charm contribution to deep inelastic (DIS) structure functions is
approximated by the leading order (LO) massive photon-gluon fusion
(`Bethe-Heitler') process \cite{BH}, since a computationally
sufficiently fast version of the next order results of \cite {LRSN},
although already announced \cite {RSN}, is not yet available. Here, we
use the LO GRV gluon density with $\Lambda^{(4)}_{LO} = 200 \mbox{ MeV}
$, and the renormalization and factorization scale is fixed at
$ 4 m_{c}^{2} $, irrespective of the value of $Q^{2} $ \cite{GRS}.
For calculating fixed-target and ISR inclusive prompt photon production
cross sections the complete NLO program of \cite{GV} is utilized, with
the photonic fragmentation functions taken from \cite{GRVf}. The small
heavy quark contribution to this process is neglected.

The construction of our distributions proceeds in two steps: first we
fix the non-singlet quark combinations $ u_{v} \equiv u - \bar{u} $,
$ d_{v} \equiv d - \bar{d} $, and $ \Delta \equiv \bar{d} - \bar{u} $
by adopting and suitably modifying the corresponding results of a recent
global fit \cite{MRSA}, then the free parameters of the valence-like
sea quark and gluon input densities are determined from data on the
proton and deuteron structure functions $F_{2}^{p,d} $ \cite
{BCDMS,SLAC,NMC} and on direct photon production in $ pp $ collisions
\cite{WA70,R806}. At a reference scale of $ Q^{2}_{0} = 4 \mbox{ GeV}
^{2} $, the non-singlet densities $ f = u_{v} $, $ d_{v} $, and
$ \Delta $ are parametrized as
\begin{equation}
  xf(x,Q^{2}_{0}) = A_{f} x^{\alpha_{f}} (1-x)^{\beta_{f}}
     (1 + \epsilon_{f} \sqrt{x} + \gamma_{f} x) \:\: .
\end{equation}
For $ \Lambda_{\overline{MS}}^{(4)} = 200 \mbox{ MeV} $ they have been
directly adopted from the MRS(A) \cite{MRSA} global fit in \cite{GRV94}.
Taking into account that the factorization scheme of \cite{GRV94}
slightly differs from the one employed in \cite{MRSA}, no readjustment
for the small $ \Lambda $-difference is necessary. For the new sets,
corresponding distributions are derived in the following way: the
contributions of $ u_{v} $, $ d_{v} $, and $ \Delta $ to $ F_{2} $ are
calculated for $ \Lambda_{\overline{MS}} ^{(4)} = 200 \mbox{ MeV} $ in
the kinematical region covered by the fixed-target data employed in the
MRS(A) analysis, and the new inputs for $\Lambda_{\overline{MS}}^{(4)}
\neq 200\mbox{ MeV} $ are fitted to these results. For the present
purpose, this procedure constitutes a sufficiently accurate substitute
for a direct fit to data.  The resulting parameters for the input (1)
are presented in Table 1.

\begin{table}[htb]
  \begin{center}
  \begin{tabular}{|c||c|c|c|c|c|}\hline
       &       &       &       &       &        \\[-0.4cm]
  $\Lambda^{(4)}/ \mbox{MeV} $
       &  150  &  250  &  300  &  350  &  400   \\[0.1cm] \hline
       &       &       &       &       &        \\[-0.4cm]
  $\Lambda^{(3)}/ \mbox{MeV} $
       &  192  &  301  &  353  &  404  &  453   \\
  $\Lambda^{(5)}/ \mbox{MeV} $
       &   95  &  168  &  206  &  245  &  285   \\[0.1cm] \hline
       &       &       &       &       &        \\[-0.4cm]
  $\alpha_{S}(M_{Z}) $
       & 0.104 & 0.113 & 0.116 & 0.119 & 0.122  \\[0.1cm]
                                                \hline \hline
       &       &       &       &       &        \\[-0.4cm]
  $ \alpha_{u}  $
       & 0.523 & 0.547 & 0.558 & 0.569 & 0.581  \\
  $ \beta_{u}   $
       & 3.991 & 3.926 & 3.895 & 3.862 & 3.829  \\
  $ \epsilon_{u}$
       & -0.30 & -0.41 & -0.47 & -0.51 & -0.55  \\
  $ \gamma_{u}  $
       &  5.33 &  5.03 &  4.92 &  4.81 &  4.68  \\[0.1cm] \hline
       &       &       &       &       &        \\[-0.4cm]
  $ \alpha_{d} $
       & 0.323 & 0.332 & 0.336 & 0.341 & 0.345  \\
  $ \beta_{d}  $
       & 4.763 & 4.664 & 4.614 & 4.563 & 4.510  \\
  $ \epsilon_{d}$
       &  4.96 &  5.25 &  5.44 &  5.58 &  5.81 \\
  $ \gamma_{d}  $
       &  6.06 &  5.44 &  5.13 &  4.78 &  4.44  \\[0.1cm] \hline
       &       &       &       &        \\[-0.4cm]
  $ A_{\Delta} $
       & 0.100 & 0.0982& 0.0974& 0.0963& 0.0952 \\
  $\alpha_{\Delta} $
       & 0.40  & 0.40  & 0.40  & 0.40  &  0.40  \\
  $\beta_{\Delta} $
       & 9.31  & 9.23  & 9.18  & 9.14  &  9.10  \\
  $\epsilon_{\Delta} $
       &  0.0  &  0.0  &  0.0  &  0.0  &   0.0  \\
  $\gamma_{\Delta} $
       & 24.5  & 25.4  & 25.7  & 26.2  &  26.8  \\[0.1cm] \hline
  \end{tabular}
  \caption{The NLO ($\overline{\mbox{MS}} $) QCD scales for $ f $
     active quarks flavours, $ \Lambda^{(f)}_{\overline{MS}} $, and
     the independent non-singlet input parameters for eq.\ (1).
     The valence quark normalizations, $ A_{u} $ and $ A_{d} $, are
     fixed by the quark number sum rules.}
  \end{center}
\end{table}

We now turn to the gluon and sea quark input distributions. Here we
first have to fix the low resolution scales $ \mu^{2} (\Lambda ) $ at
which the valence-like boundary conditions are imposed. In a LO
analysis, $ \mu^{2} $, $ Q^{2} $, and $ \Lambda $ enter the parton
density evolution and the DIS structure function $ F_{2} $  only in the
combination $ \alpha_{S}(\mu^{2}) / \alpha_{S}(Q^{2}) $. Therefore, if
the quark and gluon densities are forced to be quite similar at a
certain `central' scale $ Q_{1}^{2} $ by data, then also the inputs at
$ \mu^{2} $ have to be nearly the same irrespective of $ \Lambda $,
provided $\mu^{2} (\Lambda )$ is chosen by keeping $\alpha_{S}(\mu^{2})
 / \alpha_{S}(Q_{1}^{2}) $ constant when $ \Lambda $ is changed.
In the NLO the situation is slightly complicated by the additional
appearance of $ \Lambda $ via $ \alpha_{S}(Q^{2}) $ in the evolution
equations and the expressions for $ F_{2} $. Such effects, however, are
already incorporated in the $ \Lambda $-dependent valence distributions
derived above, suggesting that some valence moment should be kept
fixed in this case. In fact, adopting the NLO valence quark momentum
fraction $ <\! x \!>_{v} $ at $ \mu^{2} $ from \cite{GRV94}, $<\! x \!>
_{v}(\mu^{2}) = \int_{0} ^{1} \! dx \, x(u_{v}+d_{v})(x,\mu^{2}) \simeq
0.58 $, leads to very similar small-$x$ predictions in the HERA
kinematic regime for the different choices of $ \Lambda $ as illustrated
in Figure 6 below. The resulting scales $ \mu^{2}(\Lambda ) $ are listed
in Table 2. Their precise values are, of course, subject to a similar
uncertainty \cite {GRV93} as the value for $ \Lambda_{\overline{MS}}
^{(4)} = 200 $ MeV, $ \mu^{2} = 0.34 \mbox{ GeV}^{2} $ \cite {GRV94}.

\begin{table}[htb]
  \begin{center}
  \begin{tabular}{|c||c|c|c|c|c|}\hline
               &       &       &       &       &      \\[-0.4cm]
 $\Lambda^{(4)}/
   \mbox{MeV} $&  150  &  250  &  300  &  350  &  400  \\[0.1cm]
                                                  \hline \hline
               &       &       &       &       &      \\[-0.4cm]
 $\mu^{2}/\mbox
  {GeV}^{2}   $& 0.225 & 0.47  & 0.62  & 0.77  & 0.93  \\
 $<\!x\!>_{v} $& 0.579 & 0.582 & 0.579 & 0.580 & 0.582 \\[0.1cm]
                                                       \hline
 $\alpha_{g}  $&  1.9  &  1.9  &  1.9  &  1.9  &  1.9  \\
 $\beta_{g}   $&  4.0  &  4.0  &  4.0  &  4.0  &  4.0  \\[0.1cm]
                                                               \hline
               &       &       &       &       &      \\[-0.4cm]
 $2<\!x\!>_{\bar{u}
    +\bar{d}} $& 0.140 & 0.145 & 0.147 & 0.150 & 0.153 \\
 $\alpha_{\xi}$& 0.35  & 0.33  & 0.30  & 0.29  & 0.27  \\
 $\beta_{\xi} $& 7.25  & 6.84  & 6.62  & 6.42  & 6.20  \\[0.1cm]
                                                       \hline
  \end{tabular}
  \caption{The input scales $ \mu^{2} $ and the independent sea and
     gluon parameters for the valence-like input in eq.\ (2). The
     antiquark normalization is expressed via the total sea momentum
     fraction $ 2<\!x\!>_{\bar{u}+\bar{d}} $, the gluon normalization
     is fixed by the momentum sum rules.}
  \end{center}
\end{table}

At these low resolution scales $ \mu^{2} $, the gluon and antiquark
distributions are para\-met\-rized as
\begin{eqnarray}
  xg(x,\mu^{2})
        & = & A_{g} x^{\alpha_{g}} (1-x)^{\beta_{g}}      \nonumber \\
  x(\bar{u}+\bar{d})(x,\mu^{2})
        & = & A_{\xi} x^{\alpha_{\xi}} (1-x)^{\beta_{\xi}}          \\
  xs(x,\mu^{2}) & = & x\bar{s}(x,\mu^{2}) \: = \: 0 \:\: .   \nonumber
\end{eqnarray}
Our ansatz for the light sea input is slightly simpler than the one in
\cite{GRV94}, since we do not impose constraints from Drell-Yan lepton
pair production in $ pp $ interactions on the very small sea quark
densities at large $ x $. The resulting difference of the present sets
to the NLO GRV parametrization \cite{GRV94} is immaterial, except
for very special observables. Due to the energy-momentum sum rule, the
ansatz (2) consists of five independent parameters, which are inferred
from DIS structure function and direct photon production data. The sea
parameters are mainly determined by the $F_{2}^{p,d}$ results of BCDMS
\cite{BCDMS} (normalized down by $ 2.5\% $), SLAC \cite{SLAC}, and NMC
\cite{NMC}. This data is fitted in the region where these structure
functions are sensitive to the sea and gluon distributions, $ x \leq
0.3 $, and where higher-twist contributions are expected to be small,
$ Q^{2} \geq 5 \mbox{ GeV}^{2} $.
The large-$ x $ gluon density is mainly constrained by the inclusive
prompt photon production data of WA70 \cite{WA70} and R806 \cite{R806}.
Here, we fix the factorization scale at $ \mu_{F} = 0.5 \, p_{T} $,
where $ p_{T} $ denotes the photon transverse momentum, but we allow
the renormalization scale $ \mu_{R} $ to vary with $ \Lambda $. It
turns out that adopting the gluon shape parameters from the NLO ($
\Lambda _{\overline{MS}} ^{(4)} = 200 $ MeV) analysis of \cite {GRV94}
leads to a very good description of the data also for the other $
\Lambda $-values considered in this paper. We have checked that this
agreement can be only marginally improved by fitting $\alpha_{g} $ and
$ \beta_{g} $, thus their values are kept fixed. The sea and gluon
input parameters for eq.\ (2) are listed in Table 2.

A detailed comparison of our results to the data used for fixing the
valence-like inputs (2) is presented in figs.\ 1 and 2. We have selected
the three sets of distributions derived for $ \overline{\Lambda } \equiv
\Lambda_{\overline{MS}}^{(4)} \, / \, \mbox{MeV} = $ 150, 250, and 350
for the figures.
In fig.\ 1 we compare our fitted results with the BCDMS, SLAC, and NMC
fixed-target data \cite{BCDMS,SLAC,NMC} for $ F_{2}^{p} $ at $ x \leq
0.3 $ and $ Q^{2} \geq 5 \mbox{ GeV}^{2} $, the agreement with the
corresponding $ F_{2}^{d} $ results is very similar.
In fig.\ 2, the $ pp $ prompt photon data of WA70 (fixed-target) and
R806 (ISR) is compared with our distributions via the `default
quantity', $ (\sigma_{exp} - \sigma_{th}) / \sigma_{th} $, with $
\sigma_{th} $ given by the $ \overline{\Lambda } = 250 $ set. All these
comparisons exhibit an excellent agreement with the data, irrespective
of the value of $ \Lambda $.

The valence-like gluon and sea input and the valence quark densities
at $ Q^{2} = \mu^{2} $ are shown in fig.\ 3. The similarity especially
between the quark inputs for different $ \Lambda $ is striking;
it should be kept in mind, however, that $ \mu^{2} $ increases by a
factor of three when $ \Lambda_{\overline{MS}} ^{(4)} $ is raised from
150 to 350 MeV. The evolution of these inputs to $ Q^{2} $ scales of
interest for $ \alpha_{S} $ determinations is illustrated in fig.\ 4
and fig.\ 5 for the gluon distribution and the quark densities ($ u $
and $ \bar{u} $), respectively.
Finally, fig.\ 6 shows the dynamical small-$ x $ predictions for
$ F_{2}^{p} $ in comparison with recent data from the ZEUS and H1
collaborations \cite{ZEUSf2,H1f2}. Also shown are the final small-$ x $
small-$ Q^{2} $ results of E665 \cite{E665}. Within the present
experimental accuracy all sets derived in this paper are in
quantitative agreement with the HERA results; the $ \chi^{2} $ for the
149 HERA $ F_{2}^{p} $ data points lies between $ 100 $ and $ 110 $,
when the systematic and statistical errors are added quadratically and
the normalization uncertainties are neglected. Future data may impose
substantial constraints, especially at small $ Q^{2} $, $  Q^{2}
\approx 3 \mbox{ GeV}^{2} $, and very small $ x $ where the $ \Lambda
$-effects are most pronounced.

To conclude, we have derived five $\Lambda $-variants of the NLO
`dynamical' GRV parton densities, together with the standard set
\cite{GRV94} spanning the range from 0.104 to 0.122 for $\alpha_{S}
(M_{Z}) $, for use in $\alpha_{S} $ determinations from hadronic
collisions. The step size in $\alpha_{S}(M_{Z}) $ is about 0.03, which
should be sufficiently small for all applications.  It remains to be
seen whether experimental analyses will soon reach an accuracy that
renders the inclusion of $ \Lambda $-differences in the parton
distributions mandatory.
The {\sc Fortran} code for calculating these parton densities is
available via electronic mail from avogt@x4u2.desy.de.

\section*{Acknowledgement}
I thank W. Vogelsang for making available the NLO prompt photon
production computer code of \cite{GV} and for useful discussions.

\section*{Figure Captions}
\begin{description}
\item[Fig.\ 1] The proton structure function $ F_{2}^{p} $ for our
  parton density sets with $ \overline{\Lambda } \equiv \Lambda_
  {\overline{MS}}^{(4)}\, /\, \mbox{MeV} $ = 150, 250, and 350 compared
  to the data at medium $ x $ and large $ Q^{2} $ \cite{BCDMS,SLAC,NMC}
  employed in the fit of our sea and gluon inputs (2).
\item[Fig.\ 2] Comparison of the direct photon production data \cite
  {WA70,R806} used for constraining the gluon inputs in (2) with the
  results of three representative parton density sets. The factorization
  scale of the calculations is $ \mu_{F} = p_{T}/2 $. Data and curves
  have been normalized to the $ \overline{\Lambda } = 250 $ result. $
  x_{T} \equiv 2p_{T}/ \sqrt{s} $, where $ \sqrt{s} $ is the CMS energy.
\item[Fig.\ 3] The valence-like input distributions $ xf $ ($f =
  u_{v} $, $ d_{v}$, $ \bar{u} $, $ \bar{d} $, $ g $) at the small
  scales $ \mu^{2} $ given in Table 2. The strange sea $ s = \bar{s} $
  vanishes at $ Q^{2} = \mu^{2} $.
\item[Fig.\ 4] The $ Q^{2} $-evolution of our gluon distributions for
   $ \Lambda^{(4)}_{\overline{MS}} $ = 150, 250, and 350 MeV. The
   results at $ Q^{2} = 5 $ and $ 10^{4} \mbox{ GeV}^{2} $ have been
   multiplied by the numbers indicated.
\item[Fig.\ 5] The same as Fig.\ 4, but for the $ u $ and $ \bar{u} $
   quark densities. For clarity, the antiquark distributions are only
   shown where they substantially differ form the quarks.
\item[Fig.\ 6] Comparison of our dynamical leading-twist small-$x$
   predictions for $ F_{2}^{p} $ with recent ZEUS \cite{ZEUSf2} and H1
   \cite{H1f2} HERA data and low-$ Q^{2} $ E665 \cite{E665} data.
   Note that these results have not been used in our fits.
\end{description}

\newpage
\epsfig{file=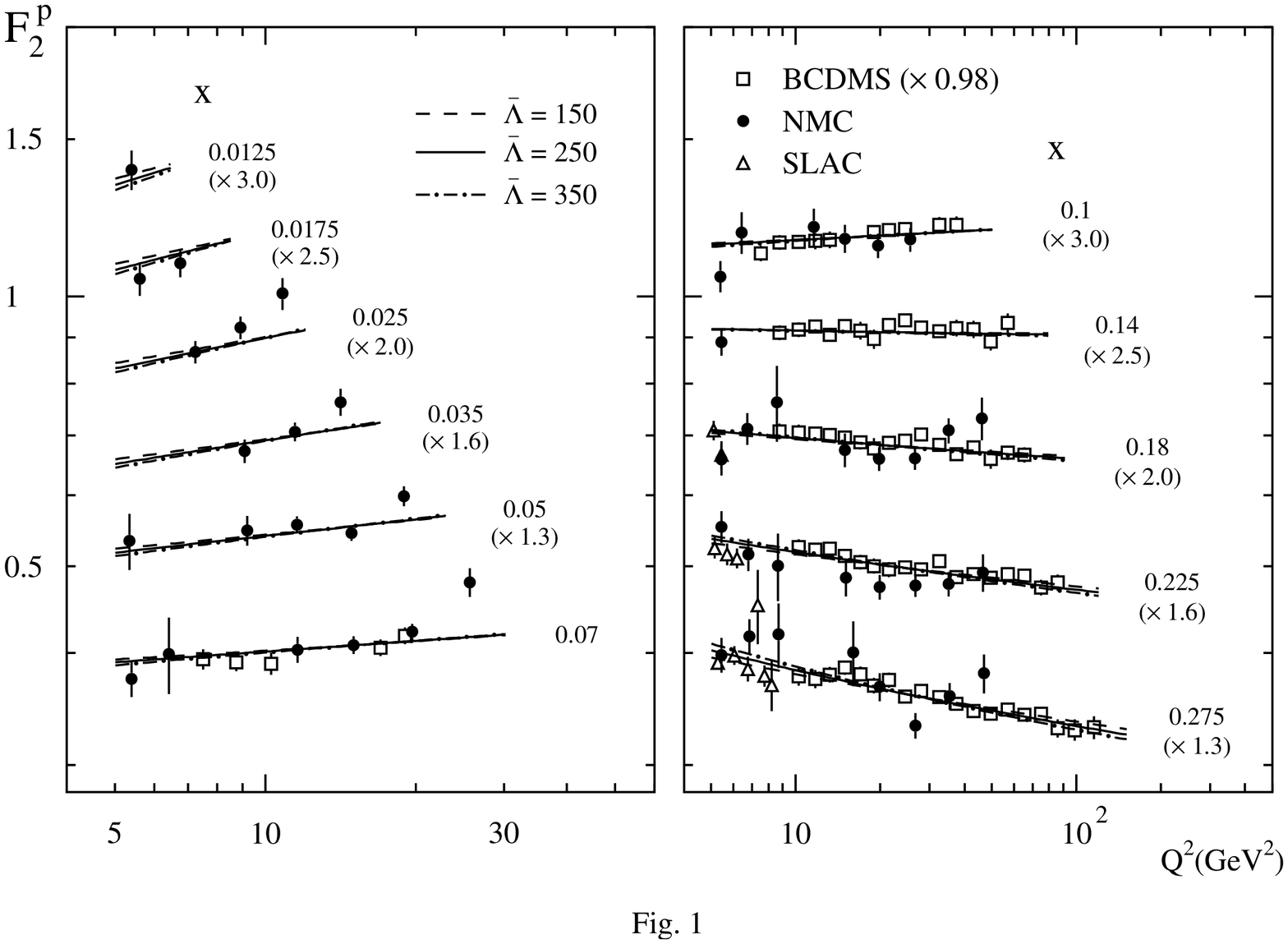,width=16cm,angle=90}

\newpage
\hspace*{0.5cm}
\epsfig{file=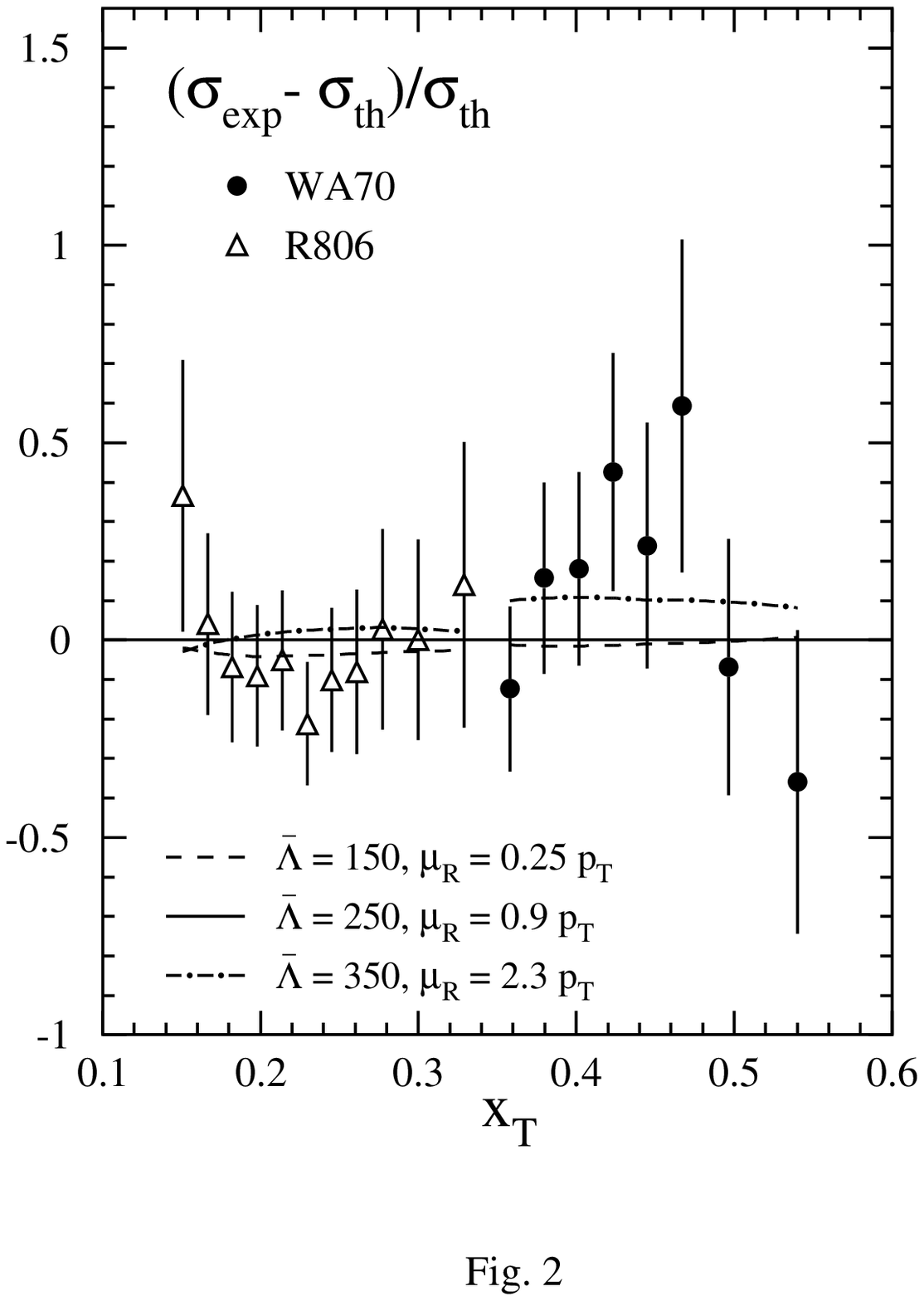,width=13cm}

\newpage
\vspace*{1cm}
\hspace*{0.5cm}
\epsfig{file=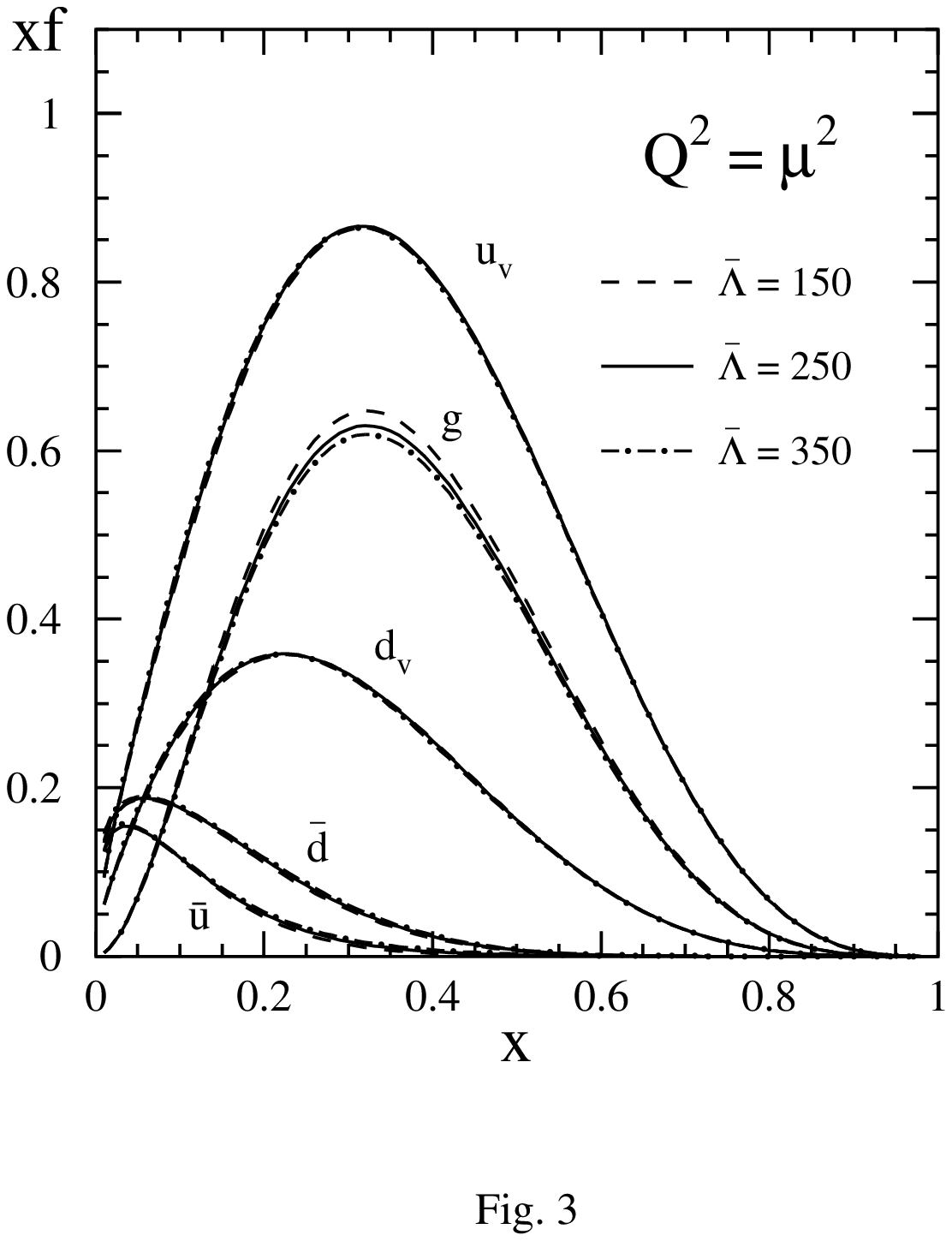,width=13cm}

\newpage
\vspace*{1cm}
\hspace*{0.5cm}
\epsfig{file=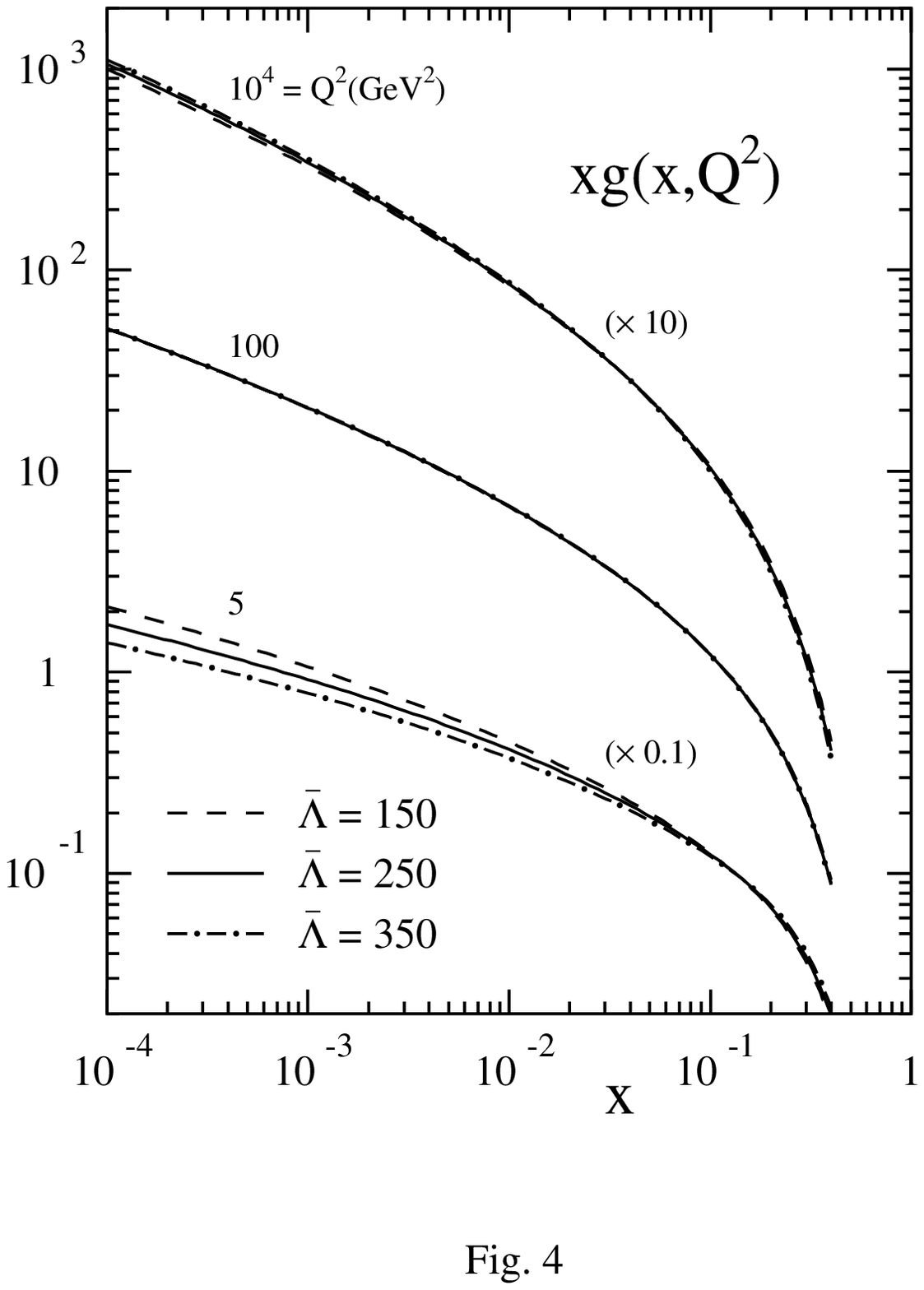,width=13cm}

\newpage
\vspace*{1cm}
\hspace*{0.5cm}
\epsfig{file=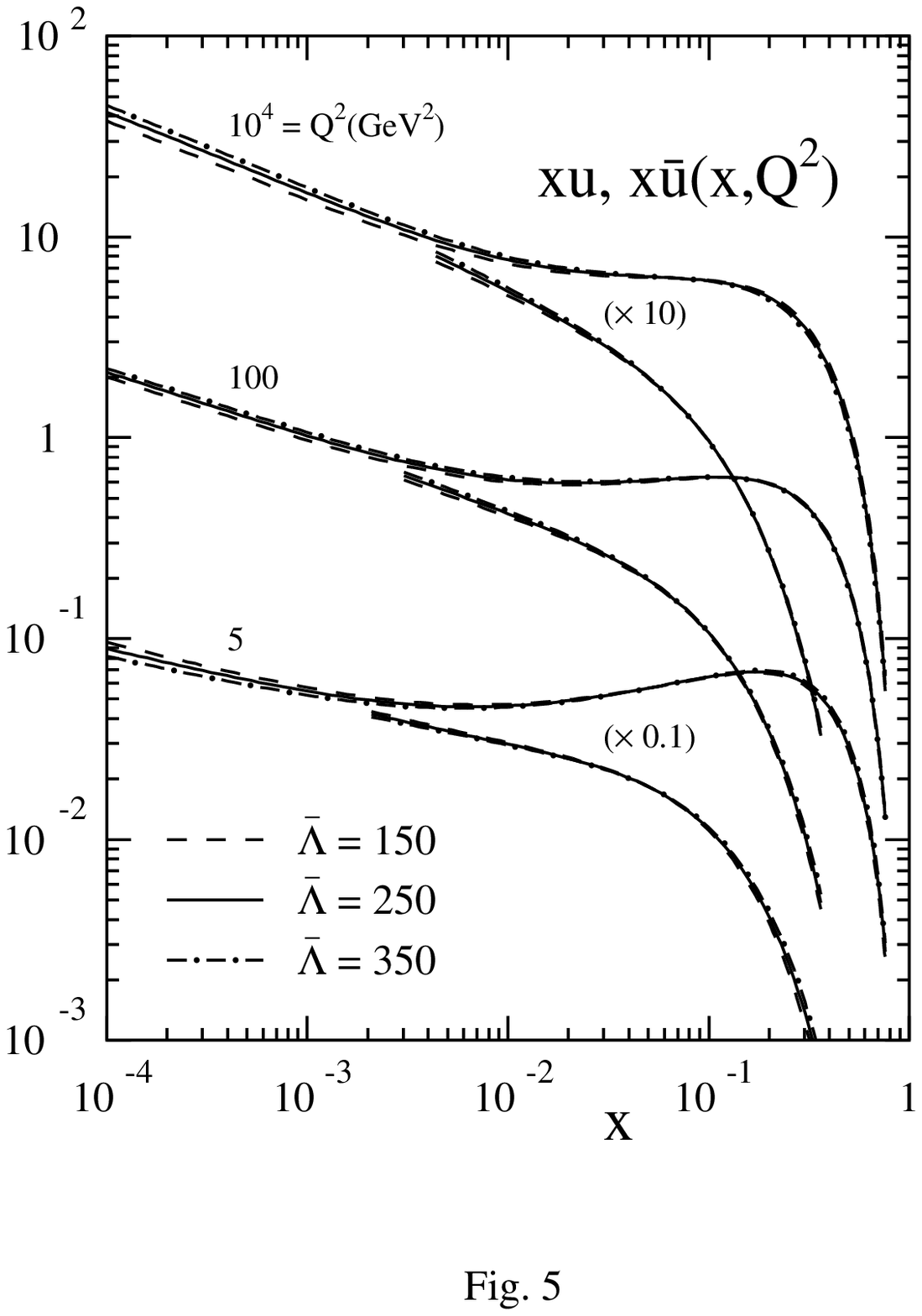,width=13cm}

\newpage
\hspace*{-1.5cm}
\vspace*{-0.5cm}
\epsfig{file=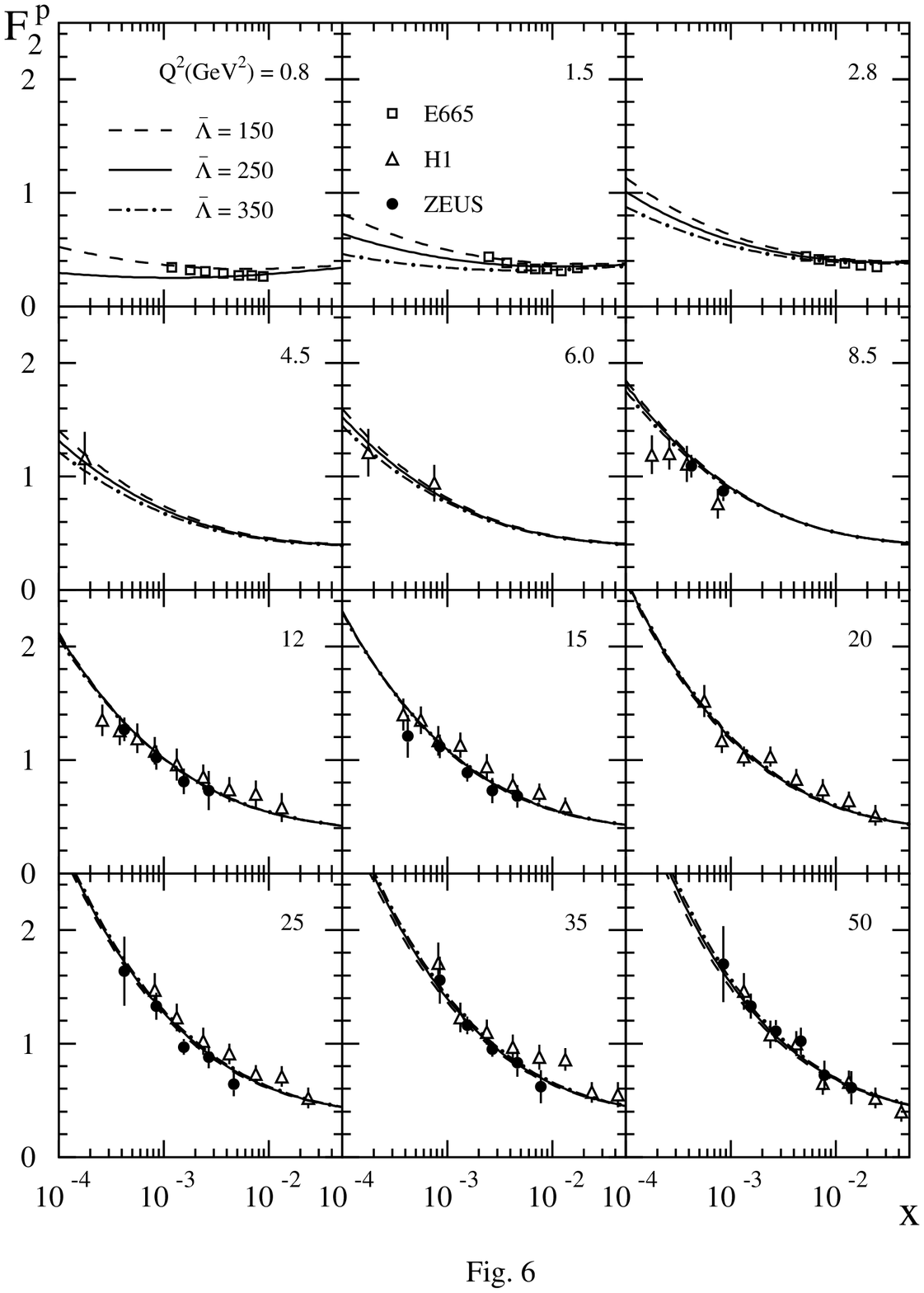,width=18cm}

\end{document}